\documentclass[aps,nofootinbib,onecolumn]{revtex4}
\usepackage{amsmath,amssymb}

\usepackage{psfrag}
\usepackage{epsfig}
\usepackage{graphicx}
\newcommand{\sgn}{\operatorname{{\mathrm sgn}}}

\begin{document}

\title{Higher-order coupled quintessence}

\author{Laura Lopez Honorez} 
\affiliation{Physics Department and IFT/CSIC, UAM, 28049 Cantoblanco, Madrid, Spain and\\
 Service de Physique Th\'eorique, ULB, 1050 Brussels, Belgium}
\affiliation{}
\author{Olga Mena} 
\affiliation{IFIC, Universidad de Valencia-CSIC, E-46071, Valencia, Spain}
\author{Grigoris Panotopoulos}
\affiliation{Departament de F\'{\i}sica Te\`orica,Universitat de
Val\`encia and IFIC, Carrer Dr. Moliner 50, E-46100 Burjassot (Val\`encia), Spain}

\date{\today}

\begin{abstract}
We study a coupled quintessence model in which the interaction with
the dark matter sector is a function of the quintessence potential. 
Such a coupling can arise from a field dependent mass term for
the dark matter field. The dynamical analysis of a 
standard quintessence potential coupled with the interaction explored here
shows that the system possesses a late time accelerated
attractor. In light of these results, we perform a fit to the
most recent Supernovae Ia, Cosmic Microwave Background and Baryon Acoustic Oscillation data sets. 
Constraints arising from weak equivalence principle violation arguments are also discussed.

\end{abstract}


\maketitle
\section{Introduction}
\label{sec:introduction}
Cosmological probes indicate that the universe we observe today possesses a flat geometry and a mass energy density made of $\sim∼30\%$ baryonic plus cold dark matter and $70\%$ dark energy, responsible for the late-time accelerated expansion. Unveiling the origin and the nature of dark energy is one of the great challenges in theoretical cosmology. The simplest candidate for dark energy is the cosmological constant, which corresponds to a perfect fluid with an equation of state $w=p/\rho=-1$. The $\Lambda$CDM model, i.e. a flat universe with a cosmological constant, is in very good agreement with current observational data. 
However, from the quantum field approach, the bare prediction for the current vacuum energy density is $\sim 120$ orders of magnitude larger than the measured value. This situation is the so-called cosmological constant problem. 
In addition, there is no proposal which explains naturally why the matter 
and the vacuum energy densities give similar contributions to the universe's energy budget at this moment in the cosmic history. This is the so-called 
\emph{why now} problem. A possible way to alleviate this problem is to assume a
time varying, dynamical fluid. The quintessence option consists on a
cosmic scalar field $\phi$ that
changes with time and varies across space, and is slowly
approaching its ground state.
 In principle, the quintessence field may couple to
the other fields, see Refs.~\cite{Damour:1990tw,Damour:1990eh, Wetterich:1994bg,Amendola:1999er,Zimdahl:2001ar,Farrar:2003uw,Das:2005yj,Zhang:2005jj,delCampo:2006vv,Bean:2007nx,Olivares:2007rt,Jackson:2009mz,Koyama:2009gd,Boehmer:2009tk}. In practice, observations strongly constrain the
couplings to ordinary matter~\cite{Carroll:1998zi}. However, interactions
within the dark sector, i.e. between dark matter and dark energy, are
still allowed. The presence of these interactions could significantly change
the universe and the density perturbations evolution, 
the latter being the seeds for
structure formation. We explore a scalar field dependent dark matter-dark energy coupling
and confront the model predictions with current cosmological data.  
For models similar to the one studied here,
see {\it e.g.}
Refs.~\cite{Damour:1990tw,Damour:1990eh,Wetterich:1994bg,Amendola:1999er,Zimdahl:2001ar,Farrar:2003uw,Das:2005yj,Zhang:2005jj,delCampo:2006vv,Bean:2007nx,Olivares:2007rt,Koyama:2009gd,Boehmer:2009tk}. The
structure of the paper is as follows. Section~\ref{sec:quint} presents
the lagrangian theory responsible for the dark sector's coupling
explored here.  Section~\ref{sec:data} describes the cosmological data sets used in the analysis. The dynamical stability of the coupled model and the fits to several cosmological observables are presented in Sec.~\ref{sec:exp}. Weak Equivalence Principle violation constraints are explored in Sec.~\ref{sec:wep}. We draw our conclusions in Sec.~\ref{sec:concl}.

\section{Coupled Quintessence}
\label{sec:quint}
Let us consider an interaction between the dark energy scalar field
$\phi$ and a cold dark matter field $\Psi$  through the dark matter mass term $m_{dm}(\phi) \bar\Psi \Psi$. This form of interaction is inspired by the 
universal coupling to all species present in scalar-tensor theories in the Einstein
frame~\cite{Damour:1990tw}. Given that, observationally,  interactions
between the dark energy field and ordinary matter are strongly constrained~\cite{Carroll:1998zi},  we will assume that the dark energy field $\phi$
does not couple to baryons. At the level of the stress-energy tensor
conservation equations, a dark energy-dark matter interaction
$m_{dm}(\phi) \bar\Psi \Psi$ implies
 \begin{eqnarray}
  \label{eq:conservDM}
\nabla_\mu T^\mu_{(dm)\nu} &=&\beta (\phi)
T^\mu_{(dm)\,\mu}\phi_{,\nu}= - \nabla_\mu T^\mu_{(de)\nu} \quad
\mbox{with }\quad  \beta(\phi)=\frac{\partial \ln m_{dm}(\phi)}{\partial \phi}~.
\label{eq:consT} 
\end{eqnarray}
 The background evolution equations read 
\begin{eqnarray}
  \ddot \phi+ 3 H \dot \phi +V_{,\phi}&=&-\beta(\phi) \rho_{dm}~;\label{eq:phiEOM}\\
\dot \rho_{dm}+3H\rho_{dm}&=&\beta(\phi)
\rho_{dm}\dot \phi~,\label{eq:DMEOM}
\label{eq:dmeq}
\end{eqnarray}
where we have used a spatially-flat Friedmann Robertson Walker metric
($ds^2=-dt^2+a^2 {\bf dx}^2$) and the dot refers to time derivative $d/dt$. 

Mostly all of the previous studies on coupled quintessence models have
assumed that the coupling $\beta (\phi)$ is a constant, see
e.g. Ref.~\cite{Wetterich:1994bg,Amendola:2003wa}. In
this paper, we consider a coupling varying with time (see also
Ref.~\cite{Amendola:2000uh,Baldi:2010vv}), given by some power of the
potential of the dark energy field, i.e. $\beta (\phi)\propto V (\phi)^n$.
For a slow-rolling quintessence field, the former assumption is
equivalent to assume a coupling proportional to some power of the dark energy
density $\beta (\phi)\propto\rho_{de}^n$. Therefore, our choice of
interaction will naturally provide a dark sector interaction $\propto
\rho_{de}\rho_{dm}$ if $n=1$. The model presented here should be understood as
a lagrangian basis for more phenomenological approaches as, for instance, 
the one presented in Ref.~\cite{Boehmer:2009tk}. We illustrate the
case of a (coupled) 
quintessence model, characterized by an exponential potential
\begin{eqnarray}
   V(\phi)&=& M^4\exp [-\alpha\phi/M_{pl}]\quad \mbox{with }\quad m_{dm}= m_{0}\exp \left[\left(V(\phi)/\rho_{cr}^0\right)^n\right]~,\label{eq:exp}
\end{eqnarray}
where $\rho_{cr}^0$ is the current critical mass-energy density today,
$H_0^2= 8\pi G/3 \rho_{cr}^0= \kappa^2/3  \rho_{cr}^0$, with
$G=1/M_{pl}^2$. The scalar field dependence on the dark matter mass
ensures that $\beta (\phi)\propto V(\phi)^n$. 
  
For the stability analysis of our coupled dark matter-dark energy
model with  the potential of Eq.~(\ref{eq:exp}), we shall focus either on 
the matter-dominated era or on the late time dark energy domination period, 
neglecting the radiation contribution. Therefore,
\begin{equation}
  \label{constr}
\Omega_{dm}+\Omega_\phi = 1\,\qquad \mbox{ with}\qquad \Omega =
\frac{\kappa^2\rho}{3H^2}\,.
\end{equation}
The baryons have been also neglected~\footnote{We neglect the presence of radiation and baryons in the stability analysis. For numerical purposes and fits to observational probes, we include both the radiation and the baryon contributions to the total mass-energy density.}. We introduce the dimensionless variables $x,y$, as in the
uncoupled case~\cite{Copeland:1997et}
\begin{align}
  x^2 = \frac{\kappa^2 \dot {\varphi}^2}{{6}H^2}\,,\qquad
  y^2 = \frac{\kappa^2 {V}}{{3}H^2}\,.
  \label{def1}
\end{align}
The positivity of the potential energy implies that $y\geq 0$. In the new variables $x$ and $y$, the equations of state are
\begin{equation}
w_\phi= \frac{p_{\phi}}{\rho_{\phi}}=\frac{x^2-y^2}{x^2+y^2}\,,\qquad w_{\text{tot}}
=\frac{p_{tot}}{\rho_{tot}}=w_\phi \Omega_\phi=
x^2-y^2\,~.
\end{equation}
The condition for a late time accelerated expansion period is still $w_{tot}<-1/3$, as in the uncoupled case.
The Hubble evolution equation can be written as
\begin{equation}
\frac{ \dot H}{H^2}=- \frac{3}{2}(1+x^2-y^2)\,.
  \label{reH}
\end{equation}
The resulting evolution equations do not allow for a two-dimensional representation of this model since $H$ can not be uniquely determined from the evolution Eqs.~(\ref{eq:phiEOM}) and (\ref{eq:DMEOM}) using exclusively the variables $x$ and $y$. Following Ref~\cite{Boehmer:2008av}, we define a third dynamical variable $z$
\begin{equation}
  z = \frac{H_0}{H+H_0}\,.
  \label{eq:z}
\end{equation}
The condition $0\leq z \leq 1$ ensures the compactness of the phase space.

\section{Cosmological data used in the analysis}
~\label{sec:data}
In this section we describe the cosmological data used in our
numerical analysis. Three different geometrical probes (Supernovae Ia (SNIa), Cosmic Microwave Background (CMB) and Baryon Acoustic Oscillations (BAO) data sets) are exploited to derive the cosmological bounds on the 
coupled quintessence model.

\subsection{The Supernova Union Compilation}
\label{sec:sn}
The Union Compilation 2~\cite{Amanullah:2010vv} consists of an update  
of the original Union compilation \cite{Kowalski:2008ez} with 557 SNIa  
after selection cuts. It includes the recent large samples of SNIa  
from the Supernova Legacy Survey and ESSENCE Survey, and the recently  
extended data set of distant supernovae observed with the Hubble Space  
Telescope (HST). In total the Union Compilation presents 557 values of  
distance moduli ($\mu$) ranging from a redshift $z$ of 0.015 up to  
$z=1.4$.
  The distance moduli, i.e. the difference between apparent and  
absolute magnitude of the objects, is given by
\begin{equation}
\mu=5\log\Big(\frac{d_L}{Mpc}\Big)+25~,
\end{equation}
\noindent
where $d_L(z)$ is the luminosity distance, $d_L(z)=c(1+z)\int_0^z  
H(z)^{-1}dz$. The $\chi^2$ function used in the analysis reads

\begin{equation}
\chi^2_{SNIa}(c_i)=\sum_{z,z^\prime}\left(\mu(c_i,z)-\mu_{obs}(z)\right)
C^{-1}_{z:z'}\left(\mu(c_i,z^\prime)-\mu_{obs}(z^\prime)\right)  
\label{chiw0},
\end{equation}
\noindent
where, $c_i$  refer to the free  
parameters of the coupled model and $C$ is the covariance matrix with  
systematics included, see \cite{Amanullah:2010vv} for details.

\subsection{CMB first acoustic peak}
\label{sec:cmb}

We exploit the CMB shift parameter $R$, since it is the least model dependent quantity extracted from the CMB power spectrum~\cite{Wang:2006ts}, i.e. it does not depend on the present value of the Hubble parameter $H_0$. The reduced distance $R$ is written as
\begin{eqnarray}
R&=&(\Omega_m H_0^2)^{1/2}\int_0^{1089} dz/H(z)~.
\label{eq:rcmb}
\end{eqnarray}
\noindent
We use the CMB shift parameter value $R=1.7\pm 0.03$, as derived in Ref.~\cite{Wang:2006ts}, where it has been explicitly shown that the value of the shift parameter $R$ is mostly independent of the assumptions made about dark energy.
The $\chi^2$ is defined as $\chi^2_{CMB}(c_i) =[(R(c_i)-R_0)/\sigma_{R_0}]^2$.

\subsection{BAOs}
\label{sec:baos}

Independent geometrical probes are BAO measurements. Acoustic oscillations in the photon-baryon plasma are imprinted in the matter distribution.
These BAOs have been detected in the spatial
distribution of galaxies by the SDSS~\cite{Eisenstein:2005su} at a redshift
$z=0.35$ and the 2dF Galaxy Redshift Survey~\cite{Percival:2007yw}
(2dFGRS) at a redshift
$z=0.2$. The oscillation pattern is characterized by a standard ruler, $s$,
whose length is the distance that the sound can travel between the Big Bang and
recombination and at which the correlation function of dark matter
(and that of galaxies, clusters) should show a peak. While future BAO data
 is expected to provide independent measurements of the Hubble rate $H(z)$ and of the angular diameter distance $D_A(z)=d_L(z)/(1+z)$ at different redshifts, current BAO data does not allow to measure them separately, so they use the spherically correlated function
\begin{eqnarray}
D_V(z)&=&\left(D^2_A(z)\frac{c z}{H(z)}\right)^{1/3}~.
\label{eq:bao1}
\end{eqnarray}
In the following, we shall focus on the SDSS BAO measurement. The SDSS team reports its BAO measurement in terms of the $A$ parameter,
\begin{eqnarray}
A(z=0.35)&\equiv& D_V(z=0.35) \frac{\sqrt{\Omega_m H^2_0}}{0.35c}~,
\label{eq:bao}
\end{eqnarray}
where $A_{SDSS}(z=0.35)=0.469\pm0.017$.
The $\chi^2$ function is defined as 
$$\chi^2_{BAO}(c_i) =[(A(c_i,z=0.35)-A_{SDSS}(z=0.35))/\sigma_{A(z=0.35)}]^2.$$

\section{Dynamical analysis and cosmological constraints}
\label{sec:exp} 
The coupling $\beta (\phi)$ of Eq.~(\ref{eq:conservDM}), for the dark matter field dependence of Eq.~(\ref{eq:exp}), gives a dynamical coupling which reads
\begin{eqnarray}
\beta(\phi)&=&- \frac{n\alpha}{M_{pl}} \left(\frac{V}{\rho_{cr}^{0}}\right)^n.\label{eq:betaexp}
\end{eqnarray}
In the next sections, we will study the impact of such a coupling in the dynamical behavior of the system as well as in the numerical analysis performed to the cosmological data sets considered here.  

\subsection{Stability analysis}
\label{sec:stability-analysis}
We study the dynamical behavior of the dark matter-dark energy
dynamical system in the matter dominated period. In terms of the $x,y$
and $z$ variables defined in Sec.~\ref{sec:quint}, Eqs.~(\ref{eq:phiEOM}),~(\ref{eq:DMEOM}) and~(\ref{reH}) read:
\begin{align}
  \label{cx'}
  x'&=-3x+\frac34 \frac{\alpha}{\sqrt{3\pi}}\,y^{2}+\frac34 \frac{\alpha}{\sqrt{3\pi}}\,y^{2n}\frac{(1-z)^{2n}}{z^{2n}}(1-x^2-y^2)+\frac{3}{2}x(1+x^2-y^2)\,,\\
  \label{cy'}
  y'&=-\alpha \frac{\sqrt{3}}{4\sqrt{\pi}}\,xy + \frac{3}{2}y(1+x^2-y^2)\,,\\
  \label{cz'}
  z'&= \frac{3}{2}z(1-z)(1+x^2-y^2)\,.
\end{align}
\begin{table}[!htb]
\begin{tabular}[t]{|c|c|c|c|c|c|c|c|}
\hline
 $x_*$ & $y_*$ & $z_*$ & Eigenvalues & $\Omega_{\phi}$ & $w_T$& Acceleration?& Existence? \\[1ex]
\hline \hline  & & & & & &  \\[-1ex]
 $0$ & $0$ & $0$ & $\frac{3}{2}\,, \frac{3}{2}\,,-\frac{3}{2}$ & 0 & 0 &No & $\forall \, \alpha$\\[1ex]
\hline   & & & & & & &   \\[-1ex]
$\pm 1$ & 0 & 0 & $3\,, 3\,,
3\mp\sqrt{\frac{3}{\pi}}\frac{\alpha}{4} $ & 1& 1&No &$\forall \, \alpha$\\[1ex]
\hline  & & & & & & &  \\[-1ex]
 $\frac{\alpha}{4\sqrt{3\pi}}$ & $\frac14 \sqrt{16 -
  \frac{\alpha^2}{3 \pi}}$  & 0 &
$\frac{\alpha^2}{16\pi}\,,
\frac{\alpha^2}{16\pi}-3\,,\sgn(-(-\alpha^2+48\pi)^n)\infty$ & 1 & $-1 +
\frac{\alpha^2}{24 \pi}$&$\alpha^2<16\pi$ & $\alpha^2< 48\pi$\\[1ex]
\hline \hline  & & & & & &  \\[-1ex]
 $0$ & $0$ & $1$ & $-\frac32,\, -\frac32, \,\frac32$ & 0 & 0 &No & $\forall \, \alpha$\\[1ex]
\hline  & & & & & & &   \\[-1ex]
$\pm 1$ & 0 & 1 & $-3,\, 3, \, 3 \mp \frac14 \alpha \sqrt{\frac{3}{\pi}}$ & 1& 1&No &$\forall \, \alpha$\\
\hline   & & & & & &  \\[-1ex]
$\frac{2\sqrt{3\pi}}{ \alpha}$
 & $\frac{2 \sqrt{3 \pi}}{\alpha}$ & 1 &
$-\frac32,\, -\frac{3 (\alpha + \sqrt{-7 \alpha^2 + 192 \pi})}{
 4 \alpha},\,\frac{3 (-\alpha + \sqrt{-7 \alpha^2 + 192 \pi})}{
  4 \alpha} $ & $\frac{24\pi}{\alpha^2}$ & 0 &No& $\alpha^2>24\pi$ \\[1ex]
\hline   & & & & & &  \\[-1ex]
$\frac{ \alpha}{4\sqrt{3\pi}}$
 & $\frac14 \sqrt{16 - \frac{\alpha^2}{3 \pi}}$ & 1 &
$-\frac{\alpha^2}{16 \pi},\, -3 + \frac{\alpha^2 }{16 \pi}, \,-3 +  \frac{\alpha^2 }{8\pi}$
&$1$ & $-1 + \frac{\alpha^2}{24\pi}$&$\alpha^2<16\pi$ & $\alpha^2< 48\pi$ \\[1ex]
\hline  
\end{tabular}
\caption{Critical points and associated eigenvalues for the exponential
  potential model of Eq.~(\ref{eq:exp}) and $n\geq 1$.} \label{tab:exp}
\end{table}
The associated critical points are presented in
Table~\ref{tab:exp}. They are independent of the value of $n$, assumed to be $n\geq 1$. Notice that our results are very similar to the ones obtained in model C of Ref.~\cite{Boehmer:2009tk}~\footnote{In Ref.~\cite{Boehmer:2009tk}, the coupling was chosen to be proportional to the Hubble rate parameter $H_0$.}. The main difference among the results presented here and those presented in Ref.~\cite{Boehmer:2009tk} lies in the eigenvalues for the $z=0$ case. The critical points for a matter dominated period followed by an accelerated expansion are however rather equivalent. 

\begin{figure}[h!]
\begin{center}
\psfrag{x}[c][c]{\large  $x$} 
\psfrag{y}[c][c]{\large  $y$} 
\hspace*{-1cm}  
\includegraphics[height=.2\textheight]{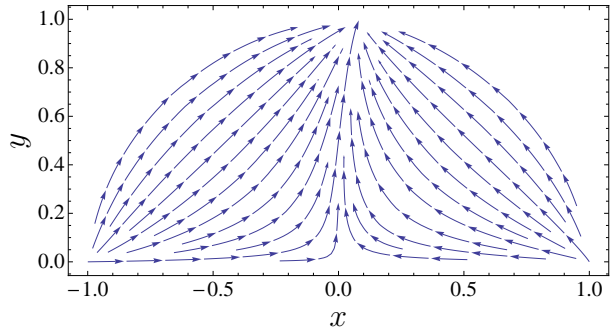}
  \caption{\it Phase space trajectories for the exponential
  potential studied in this paper at $z=1$. The plot illustrates the stable node located at
  $(x,y)=(0.16,0.99)$ for $\alpha=1$. This stable node corresponds to an 
 accelerated attractor and its existence is independent of the value of the power $n$ of the potential which appears in the dark sector interaction.}
\label{fig:PSexp}
\end{center} 
\end{figure}

The exponential coupled model of Eq.~(\ref{eq:exp}) allows for a matter dominated era at early times, corresponding to our first critical point, see Tab.~\ref{tab:exp}.  This critical point is an unstable fixed point regardless of the value of the  exponential potential parameter $\alpha$. The last critical point of Tab.~\ref{tab:exp} is an accelerated attractor for  $\alpha^2< 16\pi$. We show the phase space trajectories pointing towards this attractor in
Fig.~\ref{fig:PSexp} for $z=1$ and $\alpha=1$.
\subsection{Cosmological constraints}
\label{sec:cosm-constr-exp}
We have already shown that the background dynamics of the exponential
potential coupled model studied here offers a suitable framework to
describe the late time accelerated expansion of the Universe. We now
present the constraints which arise from the data sets described in
Sec.~\ref{sec:data}.  Both baryon and radiation contributions to the
expansion rate have been included in the following analysis. In the discussion, we make use of the individual chi-square functions, and the global chi-square is defined by
\begin{equation}
  \chi^2_{tot}(c_i)=\chi^2_{SNIa}(c_i)+\chi^2_{BAO}(c_i)+\chi^2_{CMB}(c_i)~,
\end{equation}
where $c_i$ refers to the free parameters of the coupled model under
study. The coupled model analyzed here contains three parameters $\alpha, n$
and $M$. Two parameters determine the scalar field  exponential
potential: its amplitude is set by the mass scale $M$ and 
$\alpha$ appears in the argument of the exponential. The parameter $n$ 
fixes the power of the scalar field potential appearing in the dark sector
interaction. Notice that, in this interacting quintessence model, 
$\alpha$ multiplied by $n$ plays the role of a dimensionless coupling
which sets the magnitude of the interaction in the evolution
equations, see Eq.~(\ref{eq:betaexp}). 
\begin{figure}[h!]
\begin{center}
\hspace*{-1cm}  
\begin{tabular}{cc}
\includegraphics[height=.3\textheight]{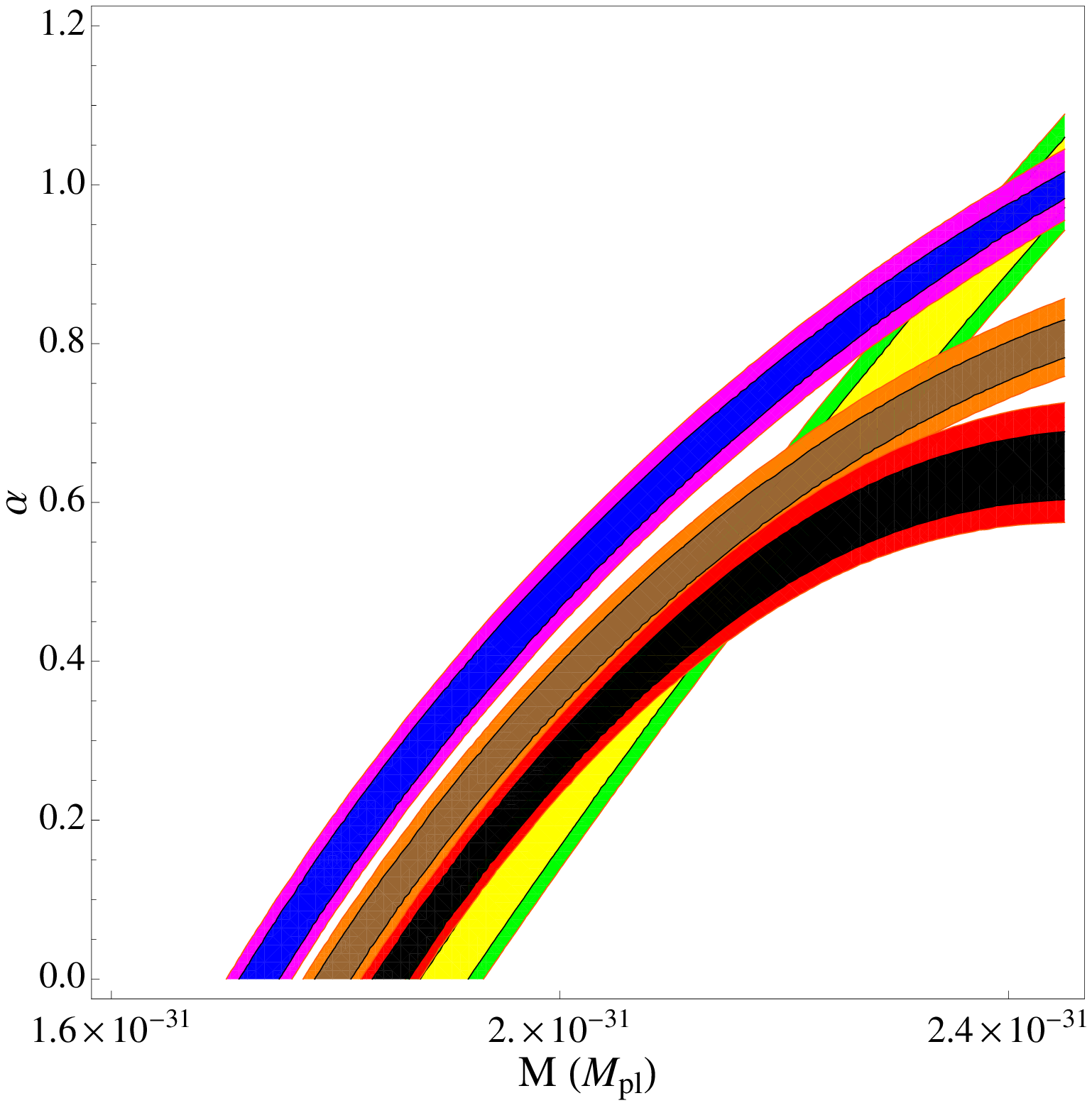}&\includegraphics[height=.3\textheight]{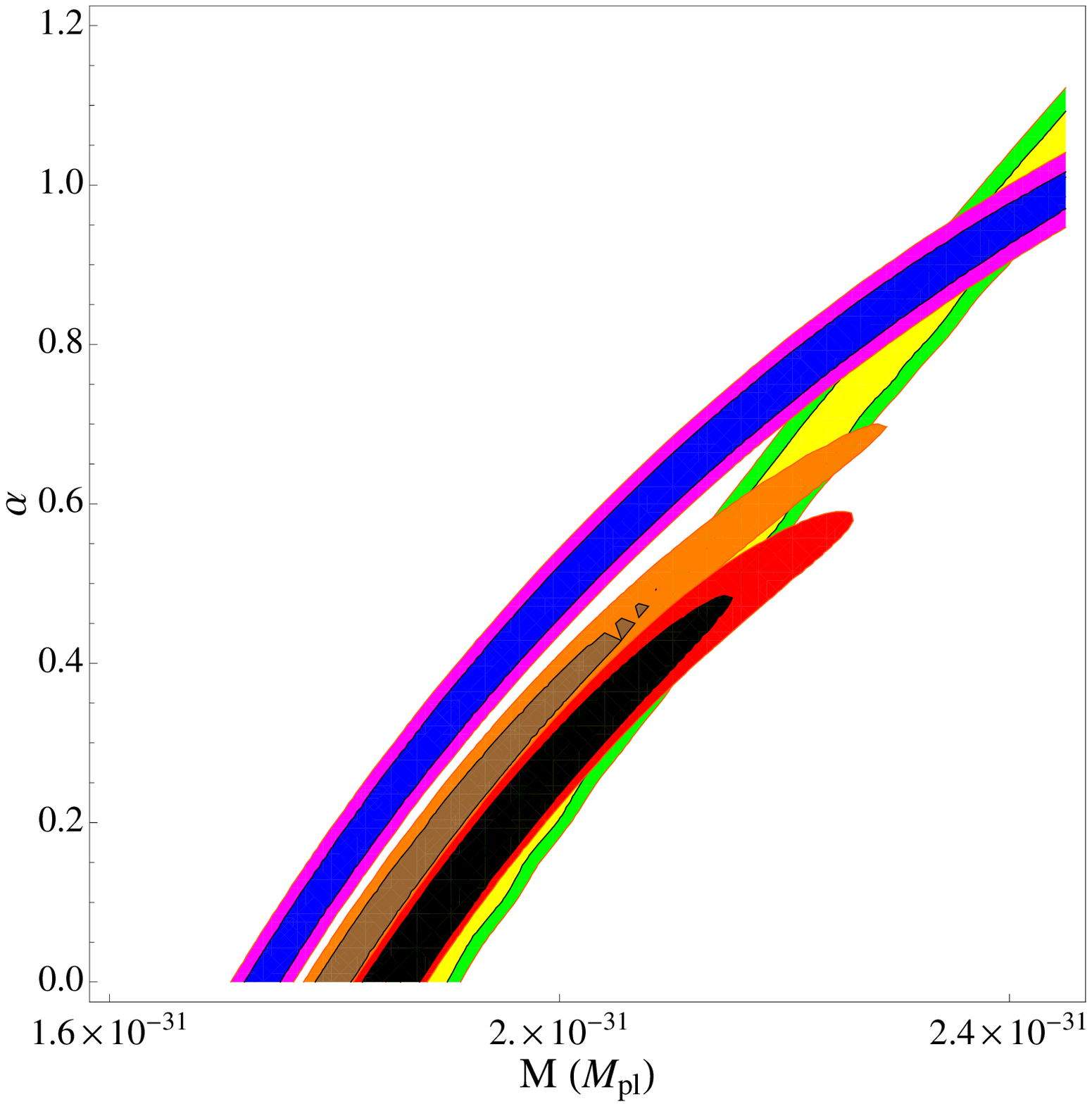}
\end{tabular}
  \caption{\it (Left panel) Analysis of the coupled exponential potential model of Eq.~(\ref{eq:exp}). The contours denote the 68.3 and 95.4 \% C.L.  allowed regions arising from a fit to SNIa data. The first three curves from left to right depict the results for $n=1,2$ and $n=4$. The last curve depicts the results arising from the fit to the uncoupled model. (Right panel) Same as in the left panel, but using the global fit results, i.e. the combined analysis to SNIa, CMB and BAO data sets.}
\label{fig:CCexp}
\end{center} 
\end{figure} 

Figure~\ref{fig:CCexp}, left panel, shows the results
of the $\chi^2$ analysis to SNIa data. In the right panel,
the results for the global fit analysis are depicted. From left to right, the curves show the $68.3$ and $95.4\%$ C.L. allowed regions for the $n=1$, $n=2$, $n=4$
and the uncoupled ($\beta(\phi)=0$) cases. For this particular analysis the 
initial conditions for the scalar field are set to $\phi_{in}=M_{pl}$ and
$\dot\phi_{in}=0$. We have checked the robustness of our results 
versus the scalar field initial conditions. Our conclusions remained 
unchanged, in agreement with the results of Ref.~\cite{Bean:2008ac}. 

In general, larger values of $\alpha$ imply a smaller scalar potential. To
compensate this effect, larger values of the amplitude of the
potential $M$ are needed. This explains the shape of the degeneracy
between $\alpha$ and $M$ in Fig.~\ref{fig:CCexp}, being these two parameters positively correlated. For small values of both $M$ and $\alpha$, for instance, $(M ,\alpha)\sim (1.8\,10^{-31}M_{pl},0.1)$, the ratio
$V/\rho_{cr}^0$ appearing in the coupling term is smaller than one. When the value of $n$ is increased from 1 to 4, the uncoupled case behavior is recovered 
due the suppression of the coupling term $(V/\rho_{cr}^0)^{n}$. Indeed, notice that the $n=4$ curve is almost superimposed to the uncoupled quintessence curve in Fig.~\ref{fig:CCexp} in the low $(M, \alpha)$ region. 

For larger values of $\alpha$ and $M$ (within the allowed regions),
the ratio $V/\rho_{cr}^0$ increases. The dark sector interaction
becomes the dominant source term for the scalar field evolution in  Eq.~(\ref{eq:phiEOM}). In addition, the dark matter energy density, proportional to $ \exp \left[\left(V(\phi)/\rho_{cr}^0\right)^n\right]$ (see
Eq.~(\ref{eq:exp})), starts to dominate the total energy density. Such a large 
contribution from the dark matter energy density is not compatible
with SNIa data. This is precisely the reason for the bending of the
curves in the left panel of Fig.~\ref{fig:CCexp}. Notice that the
shape of the curves for the coupled cases differs significantly from
the uncoupled case for larger values of $\alpha$ and $M$ (within the
allowed regions). The \emph{turn-over} of the coupled model curves
occurs at different values of the parameters $M$ and $\alpha$,
depending on the value of the $n$ parameter. Namely, for $n=4$ the
\emph{turn over} shows up at smaller values of $M$ and $\alpha$ than
those corresponding to the $n=1$ case. This is due to the fact that
the $\beta(\phi)$ term given by Eq.~(\ref{eq:phiEOM}) is proportional
to $n$ and therefore the strength of the coupling term in this region
of the $M$ and $\alpha$ parameters grows with $n$.    

Notice, from the global fit results of Fig.~\ref{fig:CCexp} (right panel), that the allowed regions for large values of $n$, $\alpha$ and $M$ become significantly smaller than those arising from a fit exclusively to SNIa data. This is due to the CMB constraint which tends to favor smaller values of $M$ in both uncoupled and coupled cases. Indeed, larger values of $M$ are associated to very small values of the dark matter energy density, which directly influences the CMB shift
parameter, see Eq.~(\ref{eq:rcmb}). 

The best fit point for the uncoupled case is located at 
$(M ,\alpha)=(2\,10^{-31}M_{pl},0.25)$ and it corresponds to
$\Omega_{de}=0.72$ and $\Omega_{dm}=0.28$, being  $\Omega_{de}$ and
$\Omega_{dm}$ the current values of the dark matter and dark energy
energy-densities, respectively. The equation of state of the dark
energy field is $w_\phi=-1$. The associated global $\chi^2$ is $562$
compared to the $557$ effective degrees of freedom in our
analysis. While for the $n=1$ coupled case the fit becomes weaker, it
improves for larger values of $n$. For the $n=4$ coupled scenario, the
best fit is associated to  a $\chi^2=542$. However, the best fit point for the $n=4$ case corresponds to a cosmological constant scenario, since it is located at $\alpha=0$ and $M=1.85\,10^{-31}M_{pl}$, with the derived cosmological parameter values $\Omega_{de}=0.67,\Omega_{dm}=0.33$ and $w_\phi=-1$. Indeed, for $\Lambda$CDM universe we obtain $\chi^2= 531$. 

In summary, the cosmological data sets considered in the current analysis favor the large $n$ regime in a region of the $(M,\alpha)$ plane where both the coupling term and the dynamics of scalar field potential are negligible. In the next section we will explore the constraints from weak equivalence principle violation arguments.

\section{Weak equivalence principle constraints}
\label{sec:wep}
Let us consider an interaction between fermionic dark matter, $\Psi$,
and a light pseudo scalar boson, $\phi$, that interacts with
the dark matter through a Yukawa coupling with strength $g$, described
by the lagrangian
\begin{eqnarray}
\mathcal{L}&=& i \bar{\Psi}\gamma_\mu\nabla^\mu\Psi -
m_\psi\bar\Psi\Psi- \frac{1}{2}\nabla_\mu\phi\nabla^\mu\phi-V(\phi) 
+g\phi\bar\Psi\Psi~,
\label{eq:lagg}
\end{eqnarray}
where $m_\psi$ is the dark matter mass (independent of the scalar field). For $g\neq0$, on scales smaller than $r_s = m_\phi^{-1}$, the Yukawa
interaction acts like a long-range `fifth' force in addition to
gravity. The effective potential felt between two dark matter
particles is
\begin{eqnarray}
V(r) = -\frac{G m_\psi^2}{r} \,
\left[1+ \alpha_{\rm Yuk} \exp\left(-\frac{r}{r_s}\right)\right]~,
\end{eqnarray}
with
\begin{eqnarray}
\alpha_{\rm Yuk}& \equiv& \frac{g^2}{4 \pi}\frac{M_{pl}^2}{ m_\psi^2}~.
\end{eqnarray}
 The authors of Ref.~\cite{Friedman:1991dj} studied the impact of such a long-range interaction on both galaxy cluster masses and dark matter growth, reporting an upper bound of $\alpha_{\rm Yuk} \leq 1.3$ (for $\alpha_{\rm Yuk}> 0$). 

Kesden and Kamionkowski (K\&K in the following)~\cite{Kesden:2006zb,Kesden:2006vz}, analyzed the consequences of Weak Equivalence Principle (WEP) violation for dark-matter on galactic scales, focusing on dark-matter dominated satellite galaxies orbiting much larger host galaxies. They concluded that models in which 
the difference among dark matter and baryonic accelerations is larger than 10\% are severely disfavoured. The relevant parameter constrained in K\&K analysis is
$\sqrt{\alpha_{\rm Yuk}}$. For reasonable models of the Sagittarius satellite galaxy tidal stream, K\&K found an upper bound of $\alpha_{\rm Yuk}< 0.04$, corresponding to 
\begin{equation}
g/m_\psi < 4.2 \times 10^{-20}~{\rm GeV}^{-1}.  
\label{eq:KK}
\end{equation}

Since the K\&K limit turns out to be the most stringent one, we
exploit it here  to set constraints on the exponential coupled model
explored along the paper. We expand the interaction term $m_{dm}(\phi)
\bar{\Psi} \Psi$, keeping only the linear terms in the scalar
field. The former approximation is possible due to the fact that 
$\alpha \phi/M_{pl}\ll 1$ within the viable regions determined in the
previous section. Therefore the linear term will clearly dominate the
dark matter-scalar field interaction. We can write Eq.~(\ref{eq:exp}) as:
\begin{eqnarray}
m_{dm}(\phi) &=& m_0 \sum_{k=0}^\infty \frac{1}{k!}\left(\frac{V(\phi)}{\rho_{cr}^0}\right)^{kn}, \\
V(\phi) & \simeq & M^4\left(1-\alpha \frac{\phi}{M_{pl}}\right)~.
\end{eqnarray}
It is straightforward to obtain an expression for both the fermion mass $m_{\psi}$ and the coupling $g$ in the Yukawa interaction term as defined in Eq.~(\ref{eq:lagg})
\begin{eqnarray}
m_{\psi}& =& m_0 \,  \exp\left[\left(\frac{M^4}{\rho_{cr}}\right)^n\right]~,\\
g&=&m_\psi \frac{n\alpha}{M_{pl}}\left(\frac{M^4}{\rho_{cr}}\right)^n~.
\end{eqnarray}
Therefore, if we express the amplitude of the potential as
$M=\lambda \times 10^{-31}M_{pl}$, the limit of Eq.~(\ref{eq:KK}) becomes
\begin{equation}
\alpha < \frac{0.1 \sqrt{8\pi}}{n\left(\frac{8\pi}{3}\frac{\lambda^4}{100}\right)^n}~.
\end{equation}
For $\lambda=1.85$, which lies in the range of the viable regions obtained in Fig.~\ref{fig:CCexp},  the former bound translates into a bound on the parameter $\alpha$ of the exponential coupled model of $\alpha<0.51, 0.26$ and $0.14$ for $n=1, 2$ and $4$, respectively. These bounds are stronger than those arising from cosmological observations and further restrict the allowed regions shown in Fig.~\ref{fig:CCexp}. In addition, WEP bounds are complementary to cosmological constraints, since they have opposite trends: while cosmological bounds are rather loose when the amplitude of the potential increases, WEP limits get much stronger. Notice however that WEP bounds have been obtained using the strongest fifth force constraint, i.e. using the K\&K limit. Mildest bounds on coupled quintessence models will arise if more conservative WEP bounds are applied.
 
\section{Conclusions}
\label{sec:concl}

We have studied a time varying-interaction among the dark matter and
the dark energy sectors. The non-minimally
coupled dark energy component is identified to a dynamical
quintessence field, and it is coupled
to the dark matter field via the dark matter mass term. The form of
the interaction has been chosen to ensure an energy exchange between
dark matter and dark energy  proportional to the product of the
dark matter energy density and the $n$th power of the scalar field
potential. For a slowly rolling scalar field and $n=1$, the model
presented here provides a
possible effective lagrangian description of pure phenomenological
quadratic interacting models such as the one studied in Ref.~\cite{Boehmer:2009tk}.

The form for the scalar field self-interacting potential is assumed to
be an exponential function of this field. The model has then been shown to possess a late time stable accelerated attractor regardless of the value of the $n$ parameter.  We have also explored the constraints on this interacting model arising from the most recent SNIa, CMB and BAO data. While the fit improves slightly when allowing for a dark matter-dark energy interaction, the best fit point lies in a region in which both the coupling and the dynamics of the field are negligible. Therefore, current data does not favor this coupled dynamical model, even if it involves more parameters than the simplest cosmological scenario, i.e. a $\Lambda$CDM universe. 
A coupling between the two dark sectors can also be constrained by
Weak Equivalence Principle violation arguments. We have derived the
constraints arising from  fifth-force searches. For the exponential
potential coupled model studied in this paper, WEP constraints are very strong and much tighter than cosmological bounds. 

\section*{Acknowledgments}

 L.~L.~H was partially supported by CICYT through
the project FPA2009-09017, by CAM through the project HEPHACOS, P-ESP-00346,  by the PAU (Physics of the accelerating universe) Consolider Ingenio 2010, by the
F.N.R.S. and  the I.I.S.N.. O.~M. work is supported by the MICINN Ram\'on y Cajal contract, AYA2008-03531 and CSD2007-00060. G.~P. acknowledges financial support from FPA2008-02878, and Generalitat Valenciana under the grant PROMETEO/2008/004.
\bibliography{bibdmde-v2}

\end{document}